\documentclass[pdftex,twocolumn,epjc3]{svjour3}

\RequirePackage[T1]{fontenc}

\smartqed 

\RequirePackage{graphicx}
\RequirePackage{mathptmx}      
\RequirePackage{flushend}
\RequirePackage[numbers,sort&compress]{natbib}
\RequirePackage[colorlinks,citecolor=blue,urlcolor=blue,linkcolor=blue]{hyperref}

\RequirePackage{amsmath}
\RequirePackage{amssymb}

\RequirePackage[usenames,dvipsnames]{xcolor}
\RequirePackage{slashed}

\RequirePackage{microtype} 
\RequirePackage{bm}

\journalname{Eur. Phys. J. C}

\newcommand{\vphi}{\varphi}
\newcommand{\eps}{\varepsilon}

\newcommand{\mbf}[1]{\mathbf{#1}}
\newcommand{\trm}[1]{\textrm{#1}}
\newcommand{\tsf}[1]{\textsf{#1}}

\newcommand{\be}{\begin{equation}}
\newcommand{\ee}{\end{equation}}
\newcommand{\bea}{\begin{eqnarray}}
\newcommand{\eea}{\end{eqnarray}}
\newcommand{\bi}{\begin{itemize}}
\newcommand{\ei}{\end{itemize}}

\newcommand{\Ai}{\trm{Ai}}
\newcommand{\Gi}{\trm{Gi}}

\newcommand{\nn}{\nonumber}

\newcommand{\ket}[1]{|#1\rangle}

\newcommand{\Braket}[3]{\langle #1|#2|#3\rangle}
\renewcommand{\epsilon}{e}
\newcommand{\av}[1]{\langle#1\rangle}

\newcommand{\figref}[1]{Fig. \ref{#1}}

\newcommand{\eqnref}[1]{Eq. (\ref{#1})}
\newcommand{\eqnrefs}[2]{Eqs. (\ref{#1}) and (\ref{#2})}

\newcommand{\vkap}{\varkappa}

\begin{document}

\title{Strong field vacuum birefringence in plane wave pulses}

\author{B.King\thanksref{e1,addr1,addr2}
\and 
T. Heinzl\thanksref{e2,addr1}
\and
T. G. Blackburn\thanksref{e3,addr3}}

\thankstext{e1}{e-mail: b.king@plymouth.ac.uk}
\thankstext{e2}{e-mail: theinzl@plymouth.ac.uk}
\thankstext{e3}{e-mail: tom.blackburn@physics.gu.se}

\institute{Centre for Mathematical Sciences, University of Plymouth, Plymouth, PL4 8AA, United Kingdom\label{addr1}
\and
Deutsches Elektronen-Synchrotron DESY, Notkestr. 85, 22607 Hamburg, Germany\label{addr2} 
\and
Department of Physics, University of Gothenburg, SE-41296 Gothenburg, Sweden\label{addr3}          
}

\date{Received: date / Accepted: date}

\maketitle

\begin{abstract}
By combining an adiabatic approach based on a `locally monochromatic' approximation with a local Hilbert transform, it is demonstrated how vacuum birefringence in the strong field regime can be calculated using a rate approach suitable for Monte Carlo simulation codes. Results for the flipping of the photon's polarisation (helicity) are benchmarked with evaluation of exact expressions in a circularly (linearly) polarised plane wave of finite extent. Example probabilities are given for typical experimental parameters.
\end{abstract}

\section{Introduction}
A long-predicted phenomenon of quantum electrodynamics is that a photon propagating through an intense electromagnetic field can flip its polarisation state due to interaction with the field through an intermediate electron-positron pair. The analogy is often made of a quantum vacuum that is polarisable in a similar fashion to a nonlinear optical material: polarisation flipping is then a signal of `vacuum birefringence'. If the intense polarising field is a plane wave, the intensity parameter, $\xi$, acts as the coupling of the charges to the background. At low intensities, $\xi \ll 1$, the leading contribution to vacuum birefringence is from four-photon scattering \cite{detollis64,detollis71,Ahmadiniaz:2021gsd}. In this regime, there is now evidence \cite{Mignani:2016fwz} from polarisation measurements of strongly magnetised neutron stars that vacuum birefringence has been observed, and the STAR collaboration has reported that angular modulation of the linear Breit-Wheeler pair creation yield in ultra-peripheral heavy-ion collision experiments can be interpreted as a consequence of vacuum birefringence \cite{STAR:2019wlg,Brandenburg:2022tna}. When $\xi \gtrsim O(1)$, a perturbative approach is no longer sufficient to describe vacuum birefringence and all orders of the interaction between the virtual pair and the background field must be taken into account. If the strong field parameter $\chi \ll 1$, an effective approach based on the Heisenberg-Euler Lagrangian \cite{euler35,heisenberg36,weisskopf36} in which the fermionic fields have been integrated out, can be used to calculate polarisation flipping. However, in the region of \emph{strong fields}, $\chi \gtrsim O(1)$, such an effective approach is no longer accurate as a significant proportion of the probability corresponds to the photon transforming into a \emph{real} electron-positron pair before annihilating back into a photon. 
This is the parameter regime of interest in the current paper; propagators in fermion loops must be replaced by those `dressed' in the background field, describing the all-order interaction with the charges, as described in the Furry picture \cite{narozhny69,baier75a,meuren13,Dinu:2013gaa,dinu14b,Torgrimsson:2020gws,Aleksandrov:2023toy}. 
Such a parameter regime may be probed by scattering experiments combining a conventionally accelerated electron beam with high intensity lasers (such as at E320 \cite{Chen:22} and LUXE \cite{Abramowicz:2021zja}), or indeed in an `all-optical' set-up using laser-wakefield acceleration at the newest generation of high-power lasers \cite{danson19}.

The standard approximation framework to describe strong-field phenomena, based on the locally constant field approximation (LCFA) \cite{nikishov64,DiPiazza:2017raw,Ilderton:2018nws,Seipt:2020diz} is also known to fail at some point in this regime; we will see this occurs at centre-of-mass energies where pairs can be created by the linear Breit-Wheeler process. In contrast, we will show that a locally monochromatic approximation (LMA) \cite{Heinzl:2020ynb} can be defined and remains accurate over the full energy spectrum. (For more background, we direct the reader to recent reviews of strong-field QED \cite{Gonoskov:2021hwf,Fedotov:2022ely,Popruzhenko:2023} and vacuum polarisation in macroscopic fields \cite{King:2015tba,Karbstein:2019oej}.)

To illustrate the challenge in deriving a local approximation to photon polarisation flipping, consider the amplitude for a flip from linear polarisation state $\ket{1}$ to $\ket{2}$ in a circularly-polarised background. If one defines photon helicity states $\ket{\pm} = (\ket{1} \pm i \ket {2})/\sqrt{2}$, and rewrites the amplitude using these states, 
the relationship follows, that:
\bea 
\tsf{S}^{\tsf{cp}}_{12} = \Braket{1}{\tsf{S}^{\tsf{cp}}}{2}=\frac{i}{2}\left[\tsf{S}^{\tsf{cp}}_{++}-\tsf{S}^{\tsf{cp}}_{--}\right]. \label{eqn:twoStateRewrite}
\eea
i.e. the real part of the amplitude for polarisation flip from $\ket{1}$ to $\ket{2}$ in a circularly polarised background is related to the imaginary part of the difference in `no flip' amplitudes for a photon in a helicity eigenstate. From the optical theorem, it follows that:
\bea 
2\,\tsf{Im}\,\tsf{S}^{\tsf{cp}}_{jj} = \tsf{P}^{\tsf{cp}}_{j\to e^{+}e^{-}}, \label{eqn:opticalTheorem}
\eea
where $\tsf{P}^{\tsf{cp}}_{j\to e^{+}e^{-}}$ is the probability of nonlinear Breit Wheeler pair creation from a photon in polarisation state $\ket{j}$. Therefore $\tsf{Re}\,\tsf{S}^{\tsf{cp}}_{12}$ can be calculated using well-established methods for pair creation e.g. the locally monochromatic approximation itself. However, calculating $\tsf{Im}\,\tsf{S}^{\tsf{cp}}_{12}$ is challenging because it involves the `off-shell' contribution (i.e. the part that does not correspond to real pair creation) 
from the electron-positron propagator and converges very slowly in the transverse momentum integral (it can be done analytically in the locally constant case). Here we will demonstrate one can use a `local' once-subtracted Hilbert transform to relate the imaginary i.e. difficult to calculate part of the amplitude to the real i.e. straightforward to calculate part:
\bea 
\tsf{Im}\,\tsf{S}^{\tsf{cp}}_{12}[\xi,\eta] = \frac{\eta}{\pi} \tsf{PV}\int d\eta' \frac{1}{\eta'} \frac{\tsf{Re}\,\tsf{S}^{\tsf{cp}}_{12}[\xi,\eta']}{\eta'-\eta}, \label{eqn:OnceHilb}
\eea 
where $\tsf{PV}$ is a principal value prescription, $\eta$ is the energy parameter to be defined and $\xi=\xi(\vphi)$ is the local intensity of the background in a way to be specified in the following. Therefore, by just using the knowledge of $\tsf{P}^{\tsf{cp}}_{j\to e^{+}e^{-}}(\xi,\eta)$, one can derive the full amplitude for polarisation flipping. Here, we extend recent work on applying the Hilbert transform within the locally constant field approximation \cite{Borysov:2022cwc}, which is insufficient for the entire parameter regime, to the locally monochromatic approach. We will assess the success of the approximation by comparing with calculations for photons scattering off plane waves of finite longitudinal extent \cite{Dinu:2013gaa,dinu14b}. In the current paper, we will apply the ideas of Hilbert transforming to these plane-wave backgrounds of finite duration and thereby extend the ideas suggested by Toll in the 1950s \cite{Toll:1952rq} for static constant crossed fields, to backgrounds more relevant for upcoming and future experiments \cite{Ilderton:2016khs,king16,Bragin:2017yau,Sangal:2021qeg,Macleod:2023asi}.

\section{Plane wave pulse background}
\begin{figure}[h!!]
\centering
\includegraphics[width=4cm]{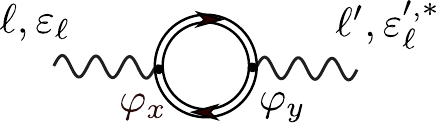}
\caption{Leading-order contribution to polarisation flipping} \label{fig:polOp1}
\end{figure}

We consider a photon of momentum $\ell$ and polarisation $\eps_{\ell}$ scattering into a state with momentum $\ell'$ and polarisation $\eps'_{\ell'}$ in a plane wave background, as illustrated in the labelled diagram in \figref{fig:polOp1}. The background is $a=eA$, where $a=a(\vphi)$, $A$ is the vector potential, $-e<0$ is the electron charge ($e>0$), and the phase $\vphi=\vkap\cdot x$ with $\vkap^{2}=0$. Due to the special kinematics in a plane wave background, which conserves transverse and lightfront momenta, if the photon remains on-shell i.e. $\ell^{2}=0$ and $\ell'^{2}=0$, it follows that $\ell=\ell'$. 
The (unrenormalised) probability can be written as:
\bea
 \tsf{P} = \Bigg|\frac{\alpha}{(4\pi\eta)^{2}}\int d\phi\,d\theta \frac{d\mbf{r}^{\perp}ds}{s(1-s)}\Theta(\theta)\tsf{T}\exp \,{i\int\limits_{\vphi_{y}}^{\vphi_{x}} \frac{\bar{\pi}^{2}}{2\vkap\cdot l}d\phi} \Bigg|^{2}, \nn \\ \label{eqn:P2}
\eea
where $\tsf{T}$ is the result of evaluating the fermion trace, $s\in[0,1]$ is the lightfront momentum fraction of the virtual positron of momentum $q$ i.e. $s=\vkap \cdot q / \vkap \cdot \ell$, the transverse momentum variable $\mbf{r}^{\perp} = [s\mbf{p}^{\perp} - (1-s)\mbf{q}^{\perp}]/m$ where $m$ is the electron mass and $\mbf{p}^{\perp}$ ($\mbf{q}^{\perp}$) is the electron (positron) momentum transverse to the wave-vector $\pmb{\vkap}$.  The phase variables are the average, $\phi = (\vphi_{x}+\vphi_{y})/2$, and difference, $\theta = \vphi_{x}-\vphi_{y}$, in phase positions of each of the vertices. The momentum appearing in the nonlinear exponent can be written $\bar{\pi}_{l}=\pi_{p}+\tilde{\pi}_{q}$ where the electron $\pi_{p}$ and positron $\tilde{\pi}_{q}$ kinetic momenta are the classical momenta solving the Lorentz equation in a plane wave, i.e.,
\bea
 \pi_{p} &=& p-a + \vkap\left(\frac{p\cdot a}{\vkap \cdot p} - \frac{a^{2}}{2\,\vkap \cdot p}\right);\nn \\
 \tilde{\pi}_{q} &=& q+a - \vkap\left(\frac{q\cdot a}{\vkap \cdot q} + \frac{a^{2}}{2\,\vkap \cdot q}\right).
\eea
We note that the nonlinear exponent in \eqnref{eqn:P2} is exactly the nonlinear Breit-Wheeler exponent in a plane-wave \cite{Seipt:2020diz}, and its complex conjugate is exactly the exponent in nonlinear one-photon pair-annihilation \cite{Tang:2019ffe}. The expression for $\tsf{T}$ is simplified if one writes photon polarisation vectors $\mbox{e}_{\ell}$ in a lightfront basis, i.e. satisfying $\vkap\cdot\mbox{e}_{\ell}=0$:
\bea 
\mbox{e}_{\ell,j} = \eps_{j} - \frac{\ell \cdot \eps_{j}}{\ell \cdot \vkap}\,\vkap. \label{eqn:photStand}
\eea 
where $j\in\{1,2\}$ and $\vkap \cdot \eps_{j}=0$. The trace term can then be written as:
\bea 
\tsf{T}&=& -\bm{\eps}\cdot \bm{\eps}^{\prime\ast}~\frac{(\mbf{a}-\mbf{a}')^{2}}{s(1-s)}\nn \\
&& +\frac{2(1-2s)^{2}}{s(1-s)}\left(\mbf{a}\cdot \bm{\eps}-s\,\mbf{r}\cdot \bm{\eps}\right)\left(\mbf{a}'\cdot \bm{\eps}^{\prime\ast}-s\,\mbf{r}\cdot \bm{\eps}^{\prime\ast}\right)\nn \\
&& - \frac{2}{s(1-s)}\left(\mbf{a}\cdot \bm{\eps}^{\prime\ast}-s\,\mbf{r}\cdot \bm{\eps}^{\prime\ast}\right)\left(\mbf{a}'\cdot \bm{\eps}-s\,\mbf{r}\cdot \bm{\eps}\right), \label{eqn:T1}
\eea where the potential evaluated at the two vertices is written as $\mbf{a}=\mbf{a}(\vphi_{x})$ and $\mbf{a}^{\prime} = \mbf{a}(\vphi_{y})$.
(A surface term of the form of the derivative of the exponent has also been removed via partial integration as is standard in plane-wave calculations.) The transverse momentum integral in $\mbf{r}^{\perp}$ can be performed analytically and the final expression must be regularised to remove the field-free divergence due to charge renormalisation. One finds the probability
\[
\tsf{P} = \bigg|\frac{\alpha}{\eta}\left[\mathcal{I}(a)-\mathcal{I}(0)\right]\bigg|^{2}, 
\] with:
\bea
\mathcal{I}(\xi)&=&  \frac{i}{8\pi}\int d\phi\,\frac{d\theta}{\theta} ds\,\Theta(\theta)\exp \, \left\{i\frac{i\theta[\mu(\theta)]}{2\eta s(1-s)}\right\} \bigg\{  \nn \\
&& \left.
-\pmb{\eps}\cdot \pmb{\eps}^{\prime\ast}~\left[\frac{(\mbf{a}-\mbf{a}^{\prime})^{2}}{s(1-s)} + \frac{8is(1-s)\eta}{\theta}\right] \right. \nn \\
&& \left. +\frac{2(1-2s)^{2}}{s(1-s)}\left(\mbf{a}\cdot \pmb{\eps}- \av{\mbf{a}}\cdot\pmb{\eps}\right)\left(\mbf{a}^{\prime}\cdot \pmb{\eps}^{\prime\ast}-\av{\mbf{a}}\cdot \pmb{\eps}^{\prime\ast}\right) \right.\nn \\
&&  - \frac{2}{s(1-s)}\left(\mbf{a}\cdot \pmb{\eps}^{\prime\ast}-\av{\mbf{a}}\cdot \pmb{\eps}^{\prime\ast}\right)\left(\mbf{a}^{\prime}\cdot \pmb{\eps}-\av{\mbf{a}}\cdot \pmb{\eps}\right)\bigg\}, \label{eqn:Ixi1}
\eea 
where the normalised Kibble mass squared is:
\[
\mu(\theta)=1-\av{\mbf{a}}^{2}+\av{\mbf{a}^{2}},
\]
and the phase window average, 
\[
\av{f} = \theta^{-1}\int_{\phi-\theta/2}^{\phi+\theta/2} f(x) dx
\]
has been used. (For the `flip' amplitude, where $\pmb{\eps}^{\prime\ast}\cdot \pmb{\eps}=0$, it was shown in \cite{Dinu:2013hsd,Dinu:2013gaa} that the lightfront momentum integral over $s$ can also be performed to acquire an even more compact expression.)

Here, we will be interested in the probability for polarisation flip, for which 
$\mathcal{I}(0)=0$. For the photon polarisation in \eqnref{eqn:photStand}, we will use 
a linear polarisation basis of the form: $\pmb{\eps}_{j} = (\delta_{j1},\delta_{j2},0)$ in the lab frame for $j \in \{1,2\}$. For the helicity basis, we choose: $\pmb{\eps}_{\pm} = (\pmb{\eps}_{1} \pm i \pmb{\eps}_{2})/\sqrt{2}$.

We refer to the `amplitude' as the quantity that occurs in the probability as a square. For example the amplitude for linear polarisation flip in a circularly polarised background, $\tsf{S}_{12}^{\tsf{cp}}$ is related to the flip probability via:
\[
\tsf{P}_{12}^{\tsf{cp}} = \left[\tsf{Re}\left(\tsf{S}_{12}^{\tsf{cp}}\right)\right]^{2}+\left[\tsf{Im}\left(\tsf{S}_{12}^{\tsf{cp}}\right)\right]^{2}.
\]
(This simple relation follows as the collision is completely elastic and the outgoing momentum integral is trivial.)

\section{Circularly polarised background}
This is an interesting background to consider, because na\"ive application of the locally constant field approximation (LCFA) and Heisenberg-Euler approaches would predict that the vacuum is \emph{not} birefringent in a circularly-polarised background, when in fact it is. This is straightforward to understand: there is a dependence on the helicity state of a photon creating a pair via nonlinear Breit-Wheeler in a circularly-polarised background, and therefore if we use the Optical Theorem from \eqnref{eqn:twoStateRewrite}, there should be a helicity dependence in photon propagation through such a background.
It is perhaps unsurprising that the LCFA and Heisenberg-Euler approaches fail in a circularly-polarised background since they are both based on constant field solutions and therefore cannot resolve how the background polarisation vector rotates with phase.
A circularly-polarised background is also relevant to experiments since it is the linear polarisation of photons that flip in this background, which is the easier polarisation state to measure in experiment (compared to helicity states). For example, the glueX experiment has measured the linear polarisation of $\sim O(10)\,\trm{GeV}$ photons to a sensitivity $\sim O(1)\%$ level using the linear trident process \cite{GlueX:2020idb}, and there have been suggestions for measuring the linear polarisation of $\trm{GeV}$ photons using pair polarimetry \cite{homma15} through the Bethe-Heitler process.

 How the LCFA fails is not completely trivial. Let us define the circularly-polarised plane wave background using the potential:
\be 
a^{\mu}(\vphi) = \frac{m\xi g(\vphi)}{\sqrt{2}}\left[\eps^{\mu}_{-}\mbox{e}^{i\vphi} + \eps^{\mu}_{+}\mbox{e}^{-i\vphi}\right],
\ee 
where $g(\vphi)$ is only non-zero for $\vphi \in [0,\Phi]$. Plugging this into \eqnref{eqn:Ixi1} for the polarisation flip $\ket{1} \to \ket{2}$, and expanding in $\theta$ in the usual way, one acquires a final integral of the form:
\[
\mathcal{I}^{\tiny\tsf{cp,lcfa}}_{12} = \int d\phi ds \frac{F(\phi)}{z}\left[\Gi'(z)+ i\Ai'(z)\right]
\]
\[
F(\phi) = \frac{\left[g^{2}(\phi)-g'^{2}(\phi)\right]\sin 2\phi + 2g(\phi)g'(\phi)\cos2\phi}{g^{2}(\phi)+g'^{2}(\phi)}
\]
\[
z =\xi \eta \sqrt{g^{2}(\phi) + g'^{2}(\phi)},
\]
where $\Ai$ and $\Gi$ are the Airy and Scorer functions respectively \cite{soares10}. The integral is not identically zero, but is very close to it in all physically interesting cases. For example, one expects $g'(\phi)/g(\phi)\sim 1/\Phi$ where $\Phi\gg1$ is the phase duration of the pulse. In the limit of $g'(\phi)/g(\phi) \to 0$, the integral is identically zero because the remaining $\phi$-dependent term, $\sin 2\phi$, is integrated over an integer number of cycles. Another way of showing the LCFA fails in a circularly-polarised background is to consider the arguments in the introduction: the probability of pair-creation from a photon in a helicity eigenstate in a circularly-polarised background is practically independent of its helicity in the LCFA (but not the LMA) \cite{Tang:2022tmn}. Therefore, by combining \eqnref{eqn:twoStateRewrite} and \eqnref{eqn:opticalTheorem} we see that the amplitude must also be effectively zero.

To show how the Heisenberg-Euler approach fails, recall that the weak-field Lagrangian can be written []:
\be 
\mathcal{L}^{\tsf{wf}} = c_{2,0}S^{2}+c_{0,2}P^{2},
\ee 
with $S = -F_{\mu\nu}F^{\mu\nu}/4$ and $P=-\widetilde{F}_{\mu\nu}F^{\mu\nu}/4$ and $c_{2,0}$, $c_{0,2}$ being constant. The scattering amplitude is then given by $\tsf{S} = -i\int d^{4}x \mathcal{L}^{\tsf{wf}}$. One finds the leading contribution to the two electromagnetic invariants is of the form:
\bea 
S &\sim& \eps \cdot \eps_{+}\, \eps^{\prime\,\ast} \cdot \eps_{-} + \eps\cdot\eps_{-}\, \eps^{\prime\,\ast} \cdot \eps_{+}; \nn \\
P &\sim & \left(\pmb{\eps} \wedge \pmb{\epsilon}_{+}\right)_{3}\,\left(\pmb{\eps}^{\prime\,\ast} \wedge \pmb{\eps}_{-}\right)_{3}+\left(\pmb{\eps} \wedge \pmb{\eps}_{-}\right)_{3}\,\left(\pmb{\eps}^{\prime\,\ast} \wedge \pmb{\epsilon}_{+}\right)_{3}
\eea
i.e. for $\eps=\eps_{1}$ and $\eps^{\prime}=\eps_{2}$, these invariants are insensitive to flipping the helicity of the background $\epsilon_{\pm} \to \epsilon_{\mp}$. Explicitly, we see each of the terms in $S$ and $P$ cancel in this case and $S=P=0$ exactly. It follows that the Heisenberg-Euler approach predicts $\tsf{S}_{12}^{\tsf{cp}}=\tsf{S}_{21}^{\tsf{cp}}=0$, which, we will see, is incorrect. (It is likely the approach can be rectified by including derivative corrections \cite{gusynin96,gusynin99,Karbstein:2021obd}.) 

\smallskip

In contrast, an LMA \emph{does} capture vacuum birefringence in circularly polarised backgrounds. This is because it includes the `fast' timescale of the carrier frequency exactly, and only locally expands the `slow' timescale carrier envelope. The LMA can be derived from \eqnref{eqn:P2} inserting \eqnref{eqn:T1} and specifying to the current case. The key point is that the nonlinear exponent, which prescribes the kinematics, can be written as 
\bea
\exp\left[i\int_{\vphi_{y}}^{\vphi_{x}} \frac{\bar{\pi}^{2}}{2\vkap\cdot l}d\phi\right] &=& \exp\left\{i\left[\frac{\theta(1+\mbf{r}^{\perp\,2}+\xi^{2}\av{g^{2}})}{2\eta s(1-s)}\right]\right\} \nn\\
&&\!\!\!\!\!\!\!\!\!\!\!\!\!\!\times\exp\left\{iz\left[\sin(\vphi_{y}+\phi_{0})-\sin(\vphi_{x}+\phi_{0})\right]\right\}\nn \\
\eea
where $\mbf{r}^{\perp} = r(\cos\phi_{0},-\sin\phi_{0})$, $z=\xi g r/[\eta s(1-s)]$ and the fast timescale is rewritten as a sum over integer harmonics using a Jacob-Anger expansion e.g.:
\[ 
\mbox{e}^{-iz\sin(\vphi+\phi_{0})} = \sum_{n=-\infty}^{\infty}J_{n}(z)\mbox{e}^{-in(\vphi-\phi_{0})},
\]
with $J_{n}(z)$ a Bessel function of the first kind. In this Fourier-transformed version, the phase integral can be performed analytically and swapped for the harmonic sum, which is often simpler to compute (more details can be found in \cite{Heinzl:2020ynb}). A direct calculation of the LMA from \eqnref{eqn:P2} faces the obstacle of the $\Theta(\theta)$ term which ensures $\vphi_{y}>\vphi_{x}$. A consequence is that the $\mbf{r}^{\perp}$ integral no longer yields a simple delta-function as it does in first-order tree-level processes, but rather a principal value part, which is slowly convergent. Instead, we use the strategy explained in the introduction combining \eqnrefs{eqn:twoStateRewrite}{eqn:opticalTheorem} with the LMA rate for photon-polarised nonlinear Breit-Wheeler pair-creation available from \cite{Tang:2022tmn,Blackburn:2023mlo}. The Hilbert transform to calculate the imaginary part of the amplitude from this is performed once, over a range of $\eta$ for fixed $\xi$ to create a reference table, which is then called and integrated locally over the pulse envelope whenever required. In other words, $\tsf{Im}\,\tsf{S}^{\tsf{cp}}_{12}[\xi,\eta]$ is a numerical function generated from a logarithmic interpolation of the Hilbert transform of the LMA for pair-creation. Here, we use the once-subtracted form from \eqnref{eqn:OnceHilb}; explicit details for the numerical method can be found in \cite{Borysov:2022cwc}.
\begin{figure}[h!!]
\centering
\includegraphics[width=8.5cm]{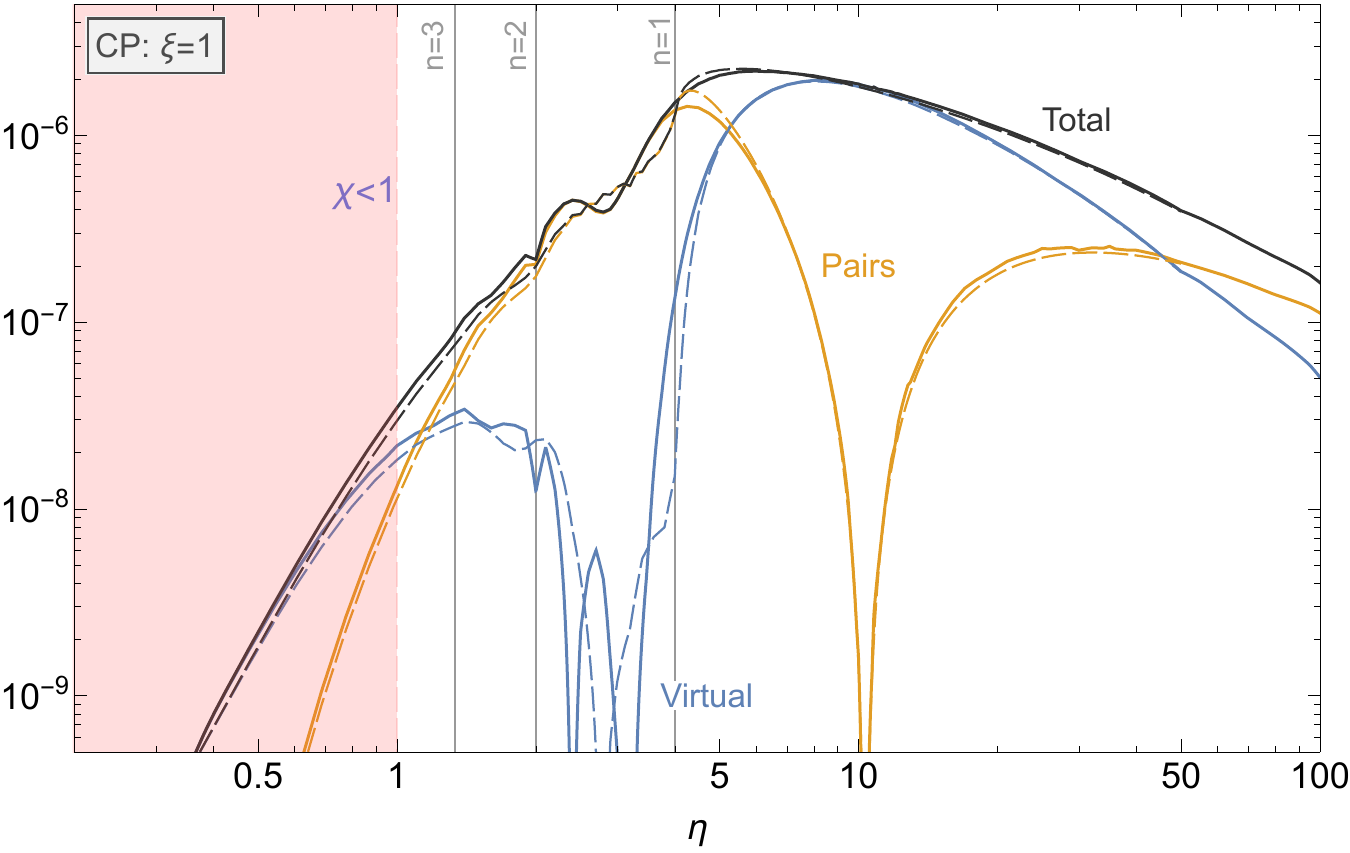}
\caption{Probability of \emph{linear polarisation} flipping in a $4$-cycle \emph{circular polarised} background with sine-squared envelope at $\xi=1$. Solid lines are the plane-wave result \eqnref{eqn:Ixi1}, dashed lines are the LMA. The edges of the harmonic ranges, $n$, for pair creation are indicated.} \label{fig:LMACP1}
\end{figure}

We compare the LMA for the probability of linear polarisation flipping in a circularly polarised background with a calculation of the direct plane-wave result in \eqnref{eqn:Ixi1}. The envelope function is chosen to be:
\bea
g(\phi) = \sin^{2}\left(\frac{\phi}{2N}\right); \quad 0<\phi<2\pi N, \label{eqn:ssquared}
\eea
and $g(\phi)=0$ otherwise. In \figref{fig:LMACP1} the predicted probability is compared for $\xi=1$ and $N=4$. We note that, although the virtual and real pair contributions vary significantly, the total curve remains rather smooth over the range of $\eta$. Although the number of cycles, $N$ is not very large, there is good agreement with the low-energy, weak-field as well as the high-energy limit, and the changing of sign of the virtual part is also well-captured by the LMA. In general, the accuracy of the `virtual part', which is generated from the Hilbert transform, is affected by the number of data points and range of energies integrated over (see e.g. \cite{Borysov:2022cwc}).

We notice from \figref{fig:LMACP1} when $\chi$ is increased above $\chi \approx 1$, the contribution from creation of real pairs becomes a significant part of the probability for polarisation flipping and the virtual part changes sign close to where pair creation is a maximum.
When the energy parameter is further increased past the point where the threshold for pair creation is already reached by a \emph{single} laser photon ($n=1$), i.e. by the \emph{linear} Breit-Wheeler process, the importance of real pair creation then falls again. If the energy parameter is raised still further, eventually pair creation becomes important again.

\section{Linearly polarised background}

The benefit of considering this background is that the plane-wave results can be compared to the LCFA and Heisenberg-Euler approximations. One can thus identify at what point they deviate from the exact result and where a more accurate approximation, such as the LMA, would be required.

We define a linearly-polarised plane-wave pulse through potential:
\be 
a^{\mu}(\vphi) = m\xi g(\vphi)\epsilon^{\mu}_{1}\cos\vphi,
\ee 
with the envelope $g(\vphi)$ given by the sine-squared shape used in the previous section in \eqnref{eqn:ssquared}.

The LCFA for polarisation flipping can be written as \cite{narozhny69,baier75a,Dinu:2013gaa}:
\bea 
\tsf{S}_{+-}^{\tsf{lp,lcfa}} = -\frac{\alpha}{\eta}\int d\phi \int_{4}^{\infty}dv\,\frac{\left[\Gi'(\bar{z}(\phi)) + i \Ai'(\bar{z}(\phi))\right]}{\bar{z}(\phi)\,v\sqrt{v(v-4)}};  \label{eqn:lcfa1}
\eea 
where $\bar{z}(\phi) = [v/\chi(\phi)]^{2/3}$, and for a potential $\mbf{a} = m\xi \mbf{f}(\phi)$ one has  $\chi(\phi) = \xi \eta [\mbf{f'}(\phi)\cdot \mbf{f'}(\phi)]^{1/2}$.

To arrive at the Heisenberg-Euler result, one need only perform a small-$\chi$ expansion of \eqnref{eqn:lcfa1}. Noting the asymptotic relation for $x \to \infty$ \cite{vallee2010airy}
\[
\Gi'(x) \sim -\frac{1}{\pi x^{2}}\sum_{n=0}^{\infty} \frac{(3n)!}{3^{n}n!}(1+3n)x^{-3n},
\]
it follows that for small $\chi$:
\bea 
\tsf{S}_{+-}^{\tsf{lp,lcfa}} \approx \frac{\alpha}{2\pi\eta}\sum_{n=0}^{\infty} \frac{(3n)!}{3^{n}n!}(1+3n)B_{2n+3,2n+3}\int d\phi \chi^{2(n+1)}(\phi),\nn\\  \label{eqn:lcfaWF}
\eea 
where $B$ is the Beta function ($B_{x,y} = \Gamma(x)\Gamma(y)/\Gamma(x+y)$ and $\Gamma$ is the gamma function \cite{olver97}). The leading-order term gives the weak-field Heisenberg-Euler (HE) result for a plane-wave background \cite{Dinu:2013gaa,king16}:
\bea 
\tsf{S}_{+-}^{\tsf{lp,HE}} =  \frac{\alpha}{60\pi\eta}\int d\phi \chi^{2}(\phi).
\eea
Note that the $n\geq 1$ terms from \eqnref{eqn:lcfaWF} cannot be reproduced by the weak-field Heisenberg-Euler approach without derivative corrections. Consider the incoming photons to be part of a beam with an intensity parameter $\xi_{p}$. Then one finds that the electromagnetic invariants are symmetric in `probe' and background parameters, e.g. $S$ has the form:
\bea
S \sim \left(\xi \frac{\vkap}{m}\right) \cdot \left(\xi_{p} \frac{\ell}{m}\right) = \chi \xi_{p}.
\eea
This is because both probe and background are plane waves so the invariants $S$ and $P$ vanish identically for each component independently and only the cross-terms of probe and background survive. To generate higher powers of $\chi$ in this approach would require higher powers of the invariants $S$ and $P$, and the incoming `probe' field would enter as a higher power, which is clearly not the case in \eqnref{eqn:lcfaWF}. This makes sense: in the plane-wave calculation the probe field is quantised and treated perturbatively, with only the leading order term appearing in the interaction with the classical background included to all orders (in $\xi$). Contrast this with the Heisenberg-Euler approach, in which the probe enters in the same way as the background $F = F_{\tsf{probe}} + F_{\tsf{background}}$ with the requirement that $F$ be slowly-varying compared to $m^{2}$.
\begin{figure}[h!!]
\centering
\includegraphics[width=8.5cm]{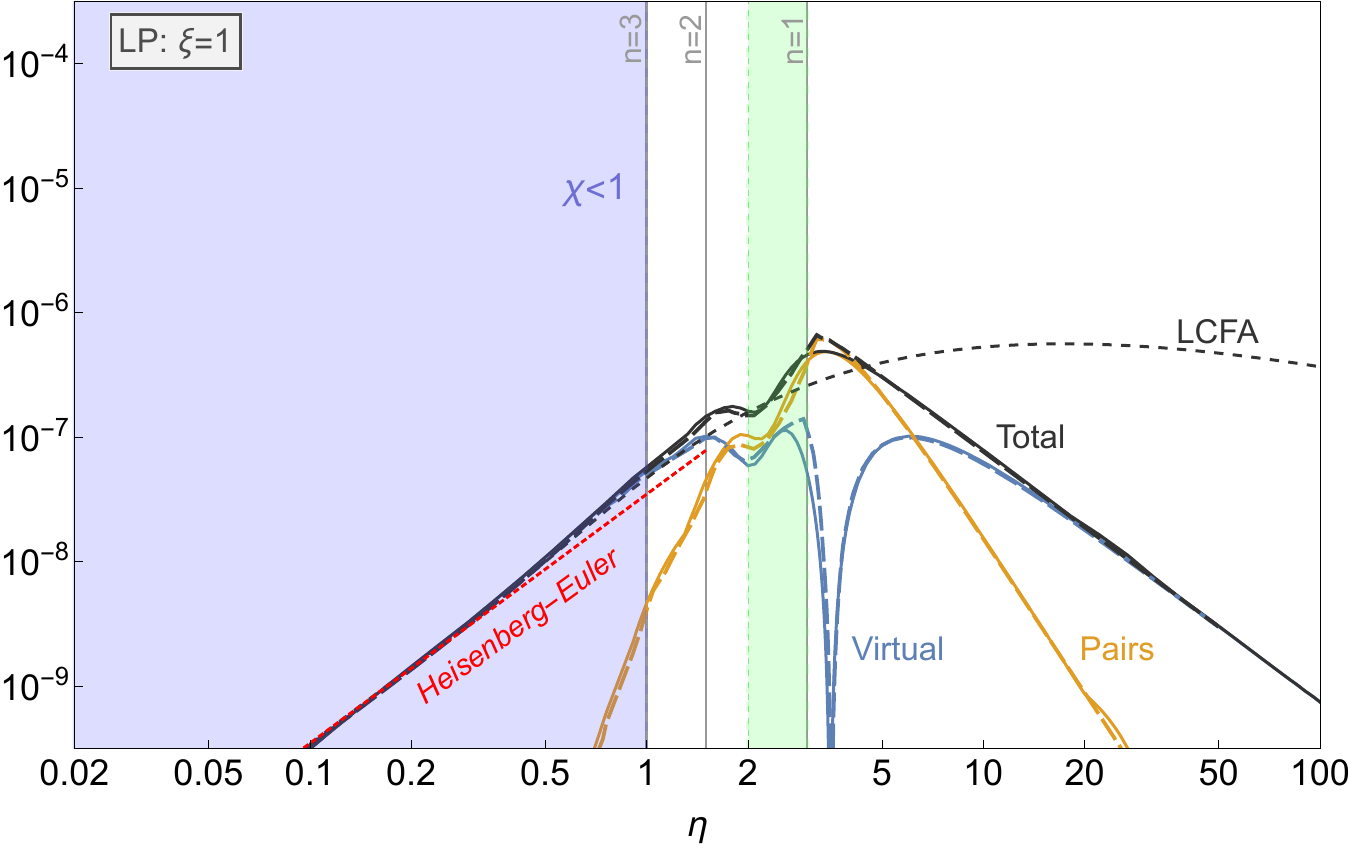}\\
\includegraphics[width=8.5cm]{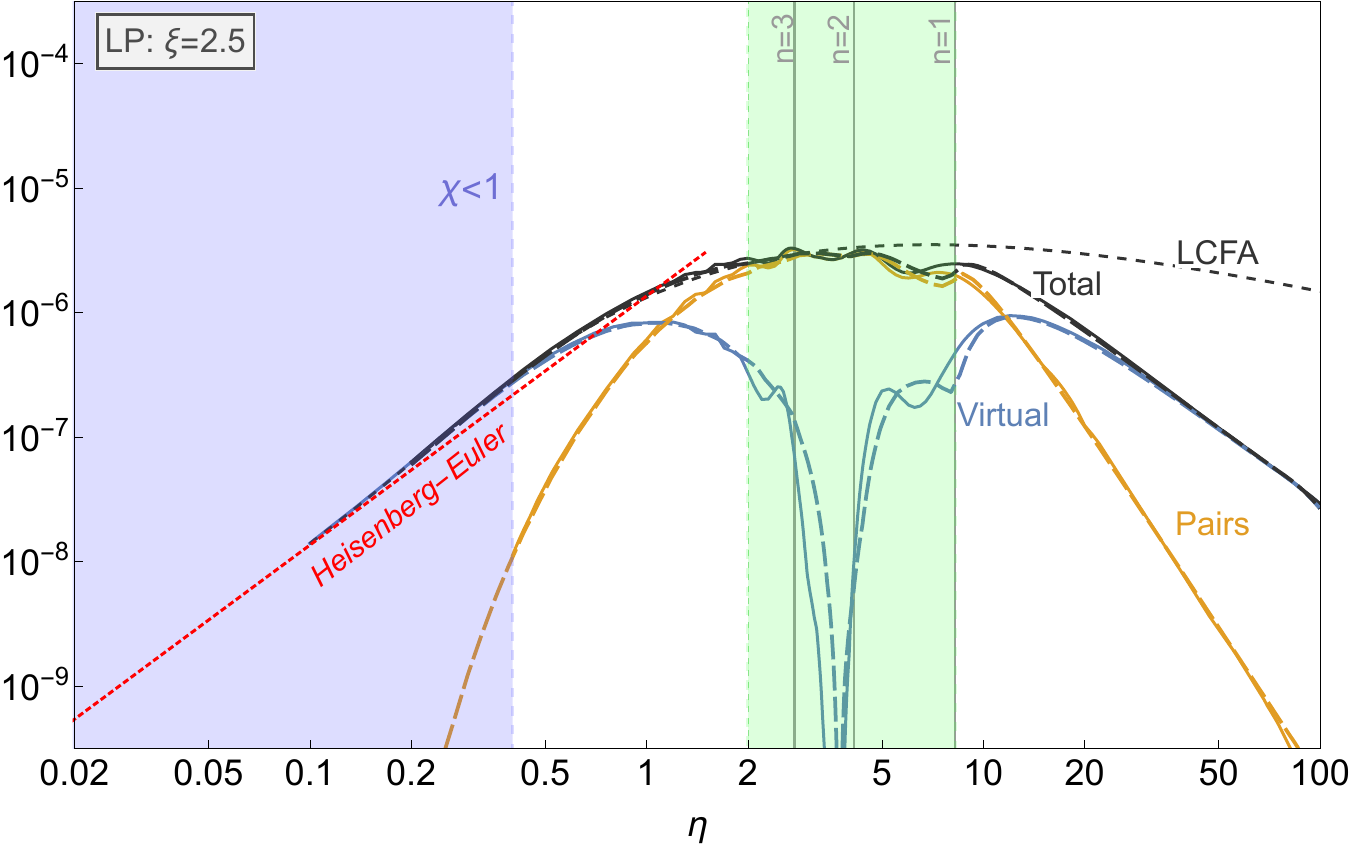}\\
\includegraphics[width=8.5cm]{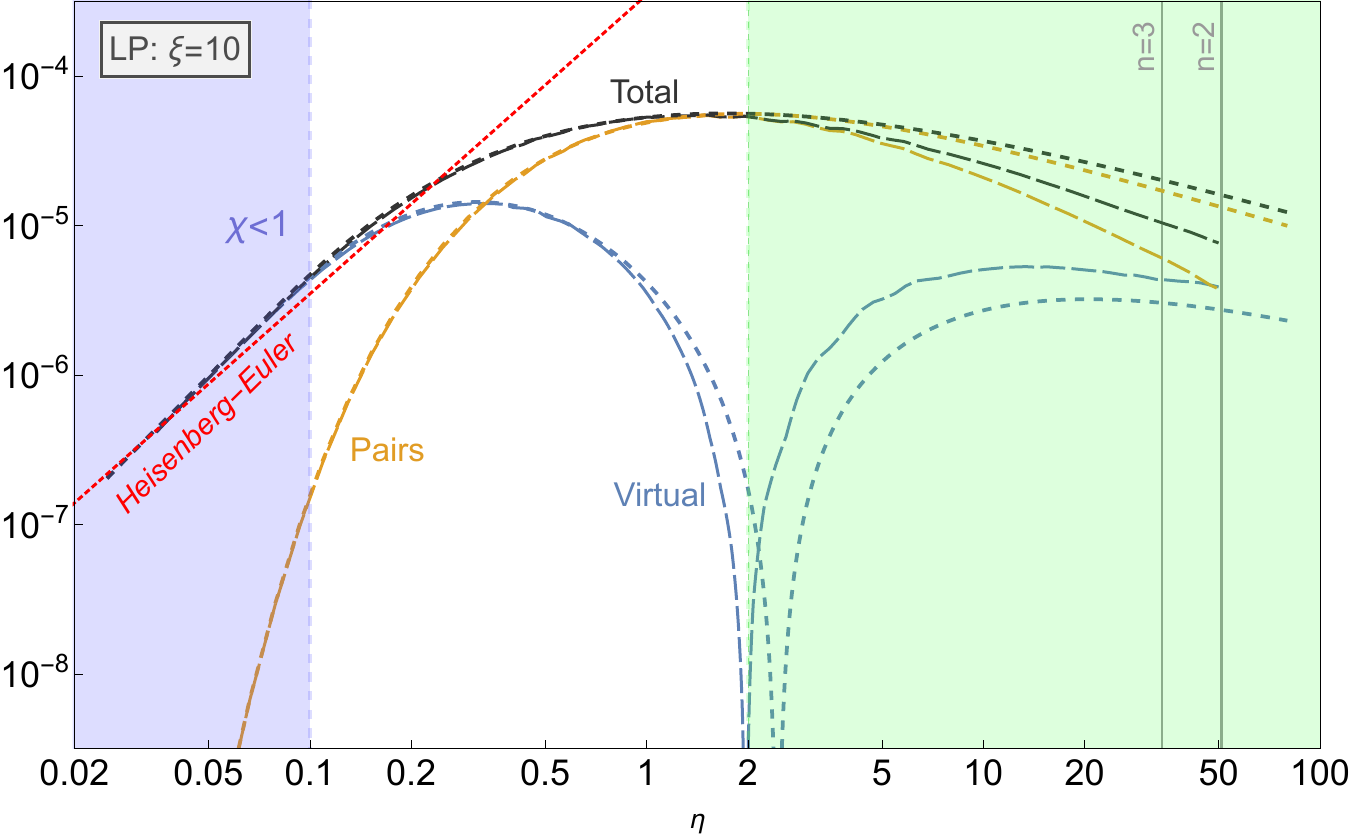}
\caption{Probability of \emph{helicity} flipping in a $4$-cycle \emph{linearly polarised} background with sine-squared envelope at $\xi=1$ (top); $\xi=2.5$ (middle) and $\xi=10$ (bottom). In the bottom plot, long-dashed lines are the LMA, and short-dashed lines are the LCFA. (Plotting scheme as in \figref{fig:LMACP1}, with the full kinematic region of \emph{linear} Breit-Wheeler, $2<\eta<2(1+\xi^{2}/2)$, being highlighted.) }  \label{fig:LMALP}
\end{figure}

We compare the LMA for the probability of helicity flipping in a linearly polarised background with a calculation of the direct plane-wave result in the top two plots of \figref{fig:LMALP} (for $\xi=1$ and $\xi=2.5$ respectively). Comparing the two figures, one sees that the weak-field Heisenberg-Euler approach agrees well with the low-$\chi$ limit of the full plane-wave result, but clearly and crucially does not include the nonlinearity associated with the strong-field interaction. The LCFA agrees with the exact result when the energy parameter is lower than the threshold for linear Breit-Wheeler. 

At energies around and above the first harmonic, the LCFA has the wrong scaling. To see this, one must take into account that the full kinematic range of linear Breit-Wheeler in a pulse, must include the fact that $\xi(\phi)$ varies between $0<\xi(\phi)<\xi$, giving the range $2< \eta < 2(1+\xi^{2})$. This failure of the LCFA is consistent with its well-known limitations in describing nonlinear Compton scattering \cite{king15d,DiPiazza:2017raw,Ilderton:2018nws} and nonlinear Breit-Wheeler pair-creation \cite{DiPiazza:2018bfu,King:2019igt} for parameters where the process can proceed by the linear, perturbative channel. The inapplicability of the LCFA for describing vacuum polarisation in the high energy limit was also pointed out in papers studying the Ritus-Narozhny conjecture at high energy \cite{Podszus:2018hnz,Ilderton:2019kqp}: here, we see at which point the LCFA begins to diverge from the exact result. In contrast, the LMA has the correct scaling in both the low- and high-energy limit, and captures oscillation of the rate around the lower harmonic positions, as well as anomalous dispersion in the virtual contribution.

Since the LMA has been benchmarked favourably with the plane-wave calculation, in the bottom plot of \figref{fig:LMALP}, we compare the LMA with the LCFA for $\xi=10$, where a direct plane-wave calculation would be numerically time-consuming. We notice that harmonic structure is pushed to very high energy parameters and that, although the LCFA agrees very well with the LMA for the contribution from real pairs, it does not exactly reproduce the sign change in the virtual part. We also notice the LCFA begins to diverge from the LMA again in the kinematic range for the linear Breit-Wheeler process, but towards end of the range, where, at this large value of $\xi$, the flip probability is actually higher than at the high-energy end of the kinematic range.

\section{Discussion}
\begin{figure}[h!!]
\centering
\includegraphics[width=8.5cm]{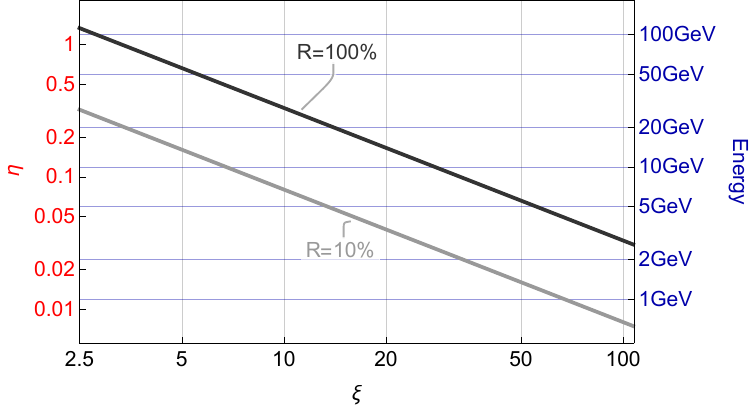}
\caption{For helicity flipping in a head-on collision with a linearly-polarised plane-wave background. $\tsf{R}$ is the ratio of the contribution to due to real pair creation compared to virtual pairs. (The curves were calculated using the LCFA.)} \label{fig:pairComp}
\end{figure}

A natural question is how to correctly include strong-field photon polarisation flipping in Monte Carlo simulations. Some approaches \cite{king16,wan2022enhanced} employ a vacuum refractive index; the real part is used to describe polarisation flip and the imaginary part is used to describe a depletion in the photon number (photon `absorption' through real pair creation). 
This approach works in the \emph{low energy} regime because although the contribution from virtual pairs to polarisation flipping is power-law suppressed, the contribution from real pair creation, because of the mass gap, is exponentially suppressed. So in this case, it is a good approximation that only virtual pairs contribute to polarisation flipping and real pairs to photon depletion. However, it is clear that \emph{both} the real and imaginary parts contribute to polarisation flip when $\chi \gtrsim 1$, and so the refractive index approach is not applicable here. It is also clear that when sufficient pairs are created, the pair plasma would generate a current that significantly modifies the photon wavefunction beyond including just a vacuum refractive index (see e.g. \cite{DiPiazza:2006gze}).
To estimate the parameters where the contribution to photon helicity flipping from real pair creation is significant, we can use the LCFA to compare the ratio at high $\xi$. In \figref{fig:pairComp} we see that at $\xi=20$, as has been achieved at e.g. Astra-Gemini \cite{cole18} and is planned for e.g. the LUXE experiment \cite{Abramowicz:2021zja}, already at $4\,\trm{GeV}$, the contribution from real pairs is 10\%, and at $15\,\trm{GeV}$ 100\% of the contribution from virtual pairs.

\begin{figure}[h!!]
\centering
\includegraphics[width=4.cm]{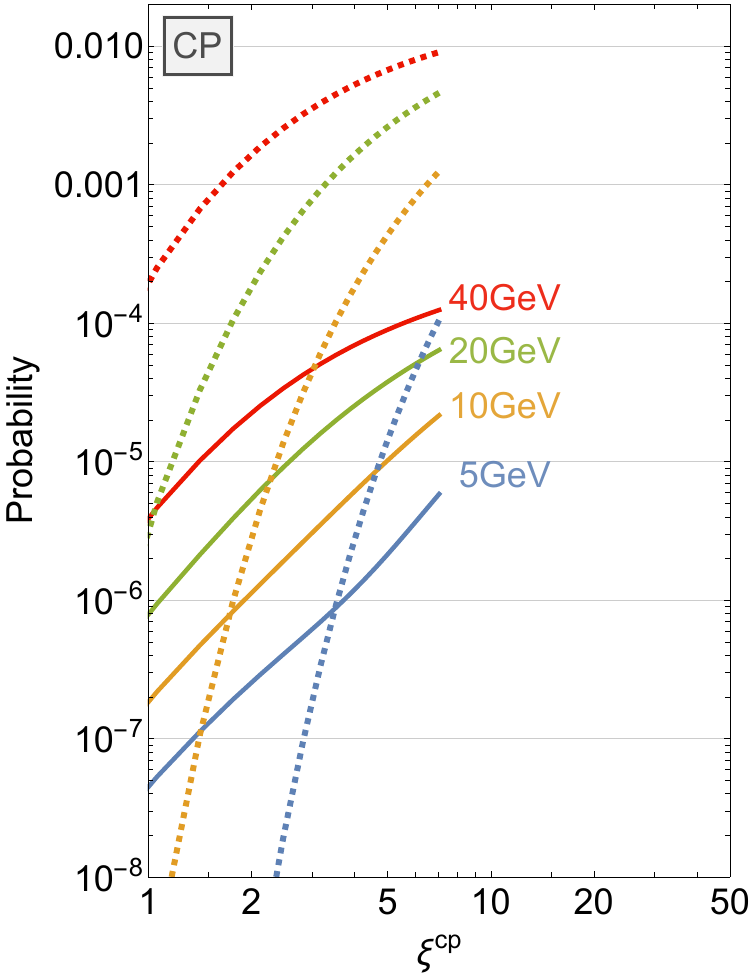}\includegraphics[width=4.cm]{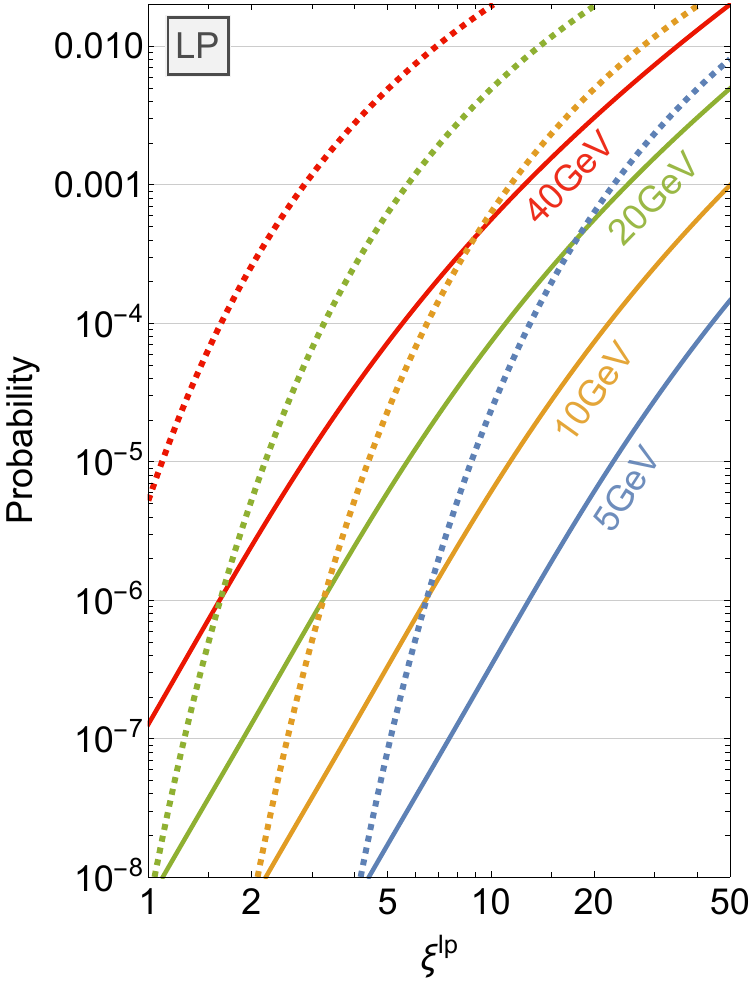}
\caption{The solid lines give the probability for a polarisation flip (left plot) or helicity flip (right plot) for a head-on collision of a photon of indicated energy with a $32$-cycle pulse of wavelength $800\,\trm{nm}~ (1.55\,\trm{eV})$ and circular polarisation (left plot) and linear polarisation (right plot). Dashed lines are the probability for pair creation.
The results in a circularly polarised background are normalised to pulse energy so that $\xi^{\tiny\tsf{cp}}=\xi^{\tiny\tsf{lp}}/\sqrt{2}$. Lines with the same colour indicate the same photon energy. Dashed lines indicate the probability for real pair creation.} \label{fig:expPlot}
\end{figure}

To experimentally verify strong field vacuum birefringence, one would ideally combine a partially polarised narrow-band source of high-energy photons with a high intensity laser and a sensitive gamma polarimeter. Examples of such photon sources include inverse Compton scattering and coherent bremsstrahlung; both of these have been suggested to be used in the LUXE experiment \cite{Abramowicz:2021zja,Borysov:2022cwc}. To give an indication of the required sensitivity for experiment, in  \figref{fig:expPlot} we plot the flip probability for a head-on collision of a photon with a circularly or linearly polarised optical plane-wave pulse at various photon energies. The relationship between the full-width-at-half-maximum pulse duration, $T$, the number of cycles $N$, and the wavelength is: $T[\trm{fs}]\approx N \lambda[\trm{nm}]/800$; in \figref{fig:expPlot}, $N=32$ and $\lambda=800\,\trm{nm}$, resulting in $T\approx32\,\trm{fs}$. For $N \gg 1$, the flipping probability scales approximately linearly with $N$.
(The four energies chosen, correspond to $\eta\in\{0.06,0.12,0.24,0.48\}$; so a $10\,\trm{GeV}$ photon colliding head-on with a frequency-doubled pulse of wavelength $\lambda=400\,\trm{nm}$, would have a flip probability on the $20\,\trm{GeV}$ line.)

One advantage of the LMA is that it predicts the correct scaling of polarisation flipping at very high energies in contrast to the LCFA. At high values of the $\chi$ parameter, based on calculations of loop processes in constant crossed fields, it has been conjectured \cite{Narozhny:1980expansion,Fedotov:2017conjecture} that the Furry expansion used in strong-field QED breaks down, and calculations must be performed to all orders in the fine-structure constant. Considering efforts to understand \cite{Podszus:2018hnz,Ilderton:2019note,Mironov:2020gbi,Heinzl:2021mji,Mironov:2022jbg,Torgrimsson:2022ndq} and conceive of experiments \cite{Blackburn:2019reaching,Yakimenko:2018kih,Baumann:2019probing,Fedeli:2021probing,Baumann:2019laser,DiPiazza:2019vwb} to reach the corresponding parameter regime of the Ritus-Narozhny conjecture of $\alpha \chi^{2/3} \approx 1$, the LMA could play a role in demarcating the parameter regime that corresponds to probabilities scaling as conjectured, from the regime where the standard QED scaling is restored.

To summarise, we verified that the tree level process acquired when cutting a `no-flip' loop (here: nonlinear Breit-Wheeler), can be used to calculate the probability of the corresponding `flip' loop process (here: photon polarisation flipping). A local monochromatic approximation for the `flip' probability can then be built from a local monochromatic approximation of the tree-level process, and its local Hilbert transform. We emphasise that the local monochromatic approximation captures linear polarisation flipping, where no other approximation scheme works (both the locally constant field approximation and an approach based on the Heisenberg-Euler Lagrangian incorrectly predict zero birefringence in a circularly-polarised bakground).
The data for local photon-polarised pair-creation rates in a linearly and circularly polarised background was generated in a range of $(\xi,\eta)$ directly using the open-source Ptarmigan \cite{ptarmigan,ptarmiganPaper} simulation code, which has been benchmarked with nonlinear Breit-Wheeler \cite{Blackburn:2021cuq} and Compton scattering \cite{Blackburn:2021rqm} calculations in a finite plane-wave background. 
Given recent interest in electron and positron spin flipping (see e.g. \cite{DelSorbo2018,Chen:2019vly,Torgrimsson:2021wcj,Torgrimsson:2021zob,Seipt2021,Wei:2022qdl,Tang:2023cyz,Seipt:2023bcw}) the analysis in the current work could also be applied to the electron mass operator.

\begin{acknowledgements}
BK acknowledges support from the Deutsche Forschungsgemeinschaft (DFG, German Research Foundation) under Germany’s Excellence Strategy – EXC 2121 ``Quantum Universe'' – 390833306.
\end{acknowledgements}

\bibliographystyle{apsrev4-1}
\bibliography{current}

\end{document}